# Correlation Time for Step Structural Fluctuations

O. Bondarchuk[*], D.B. Dougherty, M. Degawa and E.D. Williams
Department of Physics
University of Maryland
College Park, MD 20742-4111
and
M. Constantin, C. Dasgupta**, and S.Das Sarma,
Condensed Matter Theory Center
University of Maryland
College Park MD, 20742-4111

Time dependent STM has been used to evaluate step fluctuations as a function of temperature (300-450 K) on Ag(111) films grown on mica. The temporal correlation function scales as a power law in time, $t^{1/n}$ with measured values of $1/n$ varying over a range of $0.19 \pm 0.04$ to $0.29 \pm 0.04$ with no correlation on temperature. The average value of $1/n = 0.24 \pm 0.01$ is consistent with step-edge diffusion limited fluctuations ($n = z = 4$, conserved noise). The magnitude of the time correlation function and the width of the fluctuations both scale with temperature with the same apparent activation energy of $E_{eff} = 0.21 \pm 0.02$ eV, indicating that the correlation time is at most weakly temperature dependent. Direct analysis of the autocorrelation function confirms that the correlation time is at most weakly temperature dependent, and thus the apparent correlation length is strongly temperature dependent. This behavior can be reproduced by assuming that the apparent correlation length is governed by the longest wavelength of step fluctuations that can be sampled in the measurement time interval. Evaluation of the correlation time for previous measurements for Al/Si(111) ($z = 2$) yields the same conclusion about measurement time interval. In both cases the ratio of the measurement time to the effective correlation time is on the order of 10.

* Present address: Pacific Northwest National Laboratory, Richmond, WA
** Permanent Address: Department of Physics, Indian Institute of Science, Bangalore 560012, India.

**Introduction**

Direct imaging of spatial distributions and temporal variations of structures on surfaces provide the opportunity to determine thermodynamic properties from appropriately chosen correlation functions [1, 2]. Correctly interpreting the physical meaning of experimental observations of the temporal fluctuations requires careful attention to the details of experimental design. In this work, we explicitly investigate the issues of correlation length (which is often equated with system size) and measurement time with respect to the physical time constants of the system. As a test system, we have measured step fluctuations (e.g. step position vs. time) on Ag thin films over a temperature range for which the time scale of the physical fluctuations varies dramatically [2-6]. We also consider similar issues of correlation length and measurement time for fluctuations of steps on Al-terminated Si(111), for which the physical mechanism underlying the step fluctuations is different [7-9].

The results are based on an analysis of the thermodynamic correlation functions of the step fluctuations, yielding information about the physical properties governing the system. Measured correlation functions are interpreted using the Langevin description of the step fluctuations, which provides quantitative predictions for key correlation functions, dependent on whether the fluctuations are governed by a random attachment-detachment of atoms at the step edge (non-conserved noise, $n=2$), or by atomic diffusion parallel to the step edge (conserved noise, $n=4$) [1, 2, 10-14]. Key results sufficient for the analyses to be discussed here involve the time correlation function $G(t)$, which describes observations at early measurement time; the mean-squared width $w^2$, which is an average over all measurement time; and the autocorrelation function $C(t)$, which provides a useful description of the behavior at late times. The predicted forms for these functions are described in detail in the Appendix. Key results to be used in the analysis are:

$$G(t) = \langle (x(t) - x(0))^2 \rangle = \left( \frac{2\Gamma(1 - 1/n)}{\pi} \right) \left( \frac{kT}{\tilde{\beta}} \right)^{\frac{n-1}{n}} (\Gamma_n t)^{\frac{1}{n}} \qquad t \ll \tau_c \qquad (1)$$

$$w^2 = \langle (x(t) - \bar{x})^2 \rangle = \frac{kTL}{12\tilde{\beta}} \qquad (2)$$

$$C(t) = \langle (x(t)x(0)) \rangle = \frac{kTL}{12\tilde{\beta}} \exp(-t/\tau_c) \qquad t \gg \tau_c \qquad (3)$$

$$\tau_c = kT(L/2\pi)^n / \Gamma_n \tilde{\beta} \qquad (3a)$$

where $x$ is the position of the step perpendicular to the average step-edge orientation, $\Gamma$ is the gamma function, $\tilde{\beta}$ is the step stiffness, $n=2$ or 4 as discussed above, $\Gamma_n$ step mobility, $L$ is the correlation length (or system size), and $\tau_c$ is the correlation time. While our STM measurements are limited to direct observation of these one-dimensional quantities, it is important to keep in mind that the observed fluctuations are the result of the two-dimensional motion of the step-edge, which contains a range of wavelengths $\lambda$. Thus for isolated steps the time dependent correlation functions represent the integral of the contributions of all the modes of the step motion [11, 14-16]

$$G(t) = \int_{q_{min}}^{2\pi/a} G_q(t) dq; \qquad (4)$$

where $q = 2\pi/\lambda$ and $a$ is the lattice constant, e.g. the minimum distance over which the step edge position can vary. Each mode furthermore has its own fundamental time constant (decay time):

$$\tau_q = \frac{kT}{\Gamma_n \tilde{\beta} q^n}. \qquad (5)$$

As we will show in the following, the time constant associated with the different wavelengths creates a linkage between the physical interpretation of the measurements and the experimental design itself.

**<u>Experimental</u>**

Silver films of 100 nm thickness were prepared on freshly cleaved mica substrates under UHV condition (base pressure <$10^{-10}$ Torr) and imaged in an STM without exposure to air. The macor sample carrier was equipped with tungsten heater strip with clip contacts for indirect heating during film growth and also during STM imaging. Ag shots (99.9999% purity, Alfa AESAR) were evaporated from BN crucibles at deposition rate 20-40 Å/sec with the substrate held at 330-350 $^o$C [17, 18]. The deposition rate and effective film thickness were controlled by a quartz micro balance (Leybold) positioned next to the sample holder. After completing the STM measurements on each sample, the temperature-heater current relationship was measured using an alumel-chromel thermocouple, which was spot-welded to a tiny Ta tab (strip) and brought into direct contact with the film surface. Repeated sampling gave a scatter of temperatures over a range of ±20$^o$C. The reported values are the average.

To allow electrical contact to the sample area, and current stressing, wide (4mm) and thick (~3500 Å) silver leads were deposited through a mask in-situ. The sample area of 1 mm x 1mm was then evaporated through a second mask across the leads. Sample clips on the STM were used to make independent contact to the direct-current contact leads and the indirect heater current leads. Films prepared in this way could withstand an electron current up to 1 A (~$10^6$ A/cm$^2$ in the sample area).

Film purity was checked by Auger electron spectroscopy. LEED measurements indicated formation of ordered silver film with (111) orientation. Imaging of the Ag film was performed using tunneling condition of 0.6-0.8 nA and 1V and at a scan rate of ~9 pixels/ms. Under these conditions the tip-sample interactions have been shown to be negligible for a silver surface [2]. The 100nm thick films were found to have large, ~0.5-1 μm flat (111)-oriented regions with low step density as shown in Fig.1.

For observation of step fluctuation we have been using repeated STM scans across a step boundary, as shown in Fig. 2. Sampling times were 26.9 s and 39.3 s, with 512 line scans of 512 pixels each per line. Twenty such samples were measured for each set of experimental conditions. The step position $x(t)$ was extracted from each line scan by flattening the image, and then identifying the point at which surface height was midway between the heights of the upper and lower terraces. The individual $x(t)$ data sets are used to calculate individual correlation functions, and the reported correlation functions are averages over the 20 measurements. The reported error bars are the standard deviation of the average over the 20 runs. Uncertainties for the parameters determined from the data are reported as one standard deviation, and are evaluated from weighted best fits using the experimental standard deviations for weighting.

**Results**

Previous reports [2] of step fluctuations on single crystal Ag(111) samples have shown that the time correlation scales with $\sim t^{1/4}$ dependence over the temperature range of 300-450 K. The time correlation functions obtained in this work using Eq. 1 and the measured position vs. time data, $x(t)$ are shown in Fig. 3 for over a range of the temperatures investigated. The measured correlation functions show a clear power law with dynamic exponent close to 1/4 through the temperature interval from room temperature to 450 K as shown in Table I. There is no systematic variation of the exponent with temperature, and the value of the exponent averaged over the temperature range is $0.24 \pm 0.01$. The values of the pre-factor $G_1(T)$ for the time correlation function are also listed in Table 1. The mean squared widths of the step fluctuations, $w^2$, were also analyzed for all temperatures, and the values are shown in Table 1. Inspection of the temperature dependence of the mean squared width shows that it is similar to that of the magnitude of the time correlation function. This is confirmed in Figure 4, which shows that there is a linear relationship between $G_1$ and $w^2$.

The prefactor of the autocorrelation function contains the step stiffness which follows a Boltzmann temperature dependence, and the mobility which follows an Arhennius temperature dependence (see discussion section). Thus an Arhennius plot of the magnitude $G_1(T)$ of the time correlation function is a reasonable approach to extract the temperature dependence, as shown in Figure 5. Despite the large scatter, largely due to variations in temperature during measurements, the data points can be well fitted by a straight line, yielding an effective activation energy of $E_a=0.21\pm0.02$ eV. From the linear relationship of $G_1$ and $w^2$ shown in Fig. 4, the temperature dependence of the width is also governed by this effective activation energy. This interesting result is evaluated in the discussion section.

Finally, the data were also evaluated in terms of the autocorrelation function, Eq. 3. The statistics quickly become poor at the longer times needed for this analysis, and only four data sets (measured with larger sampling times) yielded autocorrelation curves with clearly interpretable features. Two of those are shown in Fig. 6, and the time constants extracted from exponential fits to the autocorrelation functions are listed in Table 1.

**Discussion**

We find that the time correlation scales as power law in time, independent of temperature, with an exponent consistent with $1/n =1/4$, as expected for step meandering governed by step-edge diffusion. The measured effective activation energy $E_a$ (Fig. 5), can be related to the kink energy ($\varepsilon_k$) and effective diffusion barrier $E_h$ along a step [2]: $E_a = 3/4\ \varepsilon_k + 1/4\ E_h$. Using $\varepsilon_k = 0.101 \pm 0.005$ eV [21], one can find $E_h = 0.54 \pm 0.08$ eV. In [2] $E_h$ was expressed as: $E_h \approx 3\varepsilon_k + E_d$, where $E_d$ is the activation barrier for diffusion along a straight step. Thus, from the measured effective activation energy we obtain $E_d \approx 0.24$ eV. This value is in reasonable agreement with calculated values of edge diffusion barriers 0.222 eV and 0.300 eV for A- and B- steps on Ag(111) respectively [22].

The magnitude of the time correlation function can be analyzed to extract information about the underlying time constant $\tau$ governing step fluctuations if the step stiffness is known. Using Eq. 1, and the relationship [1] between $\Gamma_n$ and $\tau$, $\Gamma_n = \dfrac{a^{n+1}}{\tau}$, yields:

$$\tau = \left(\frac{2\Gamma(1-1/n)}{\pi}\right)^n \left(\frac{kT}{\tilde{\beta}}\right)^{n-1} \frac{a^{n+1}}{G_1(T)^n}, \qquad (6)$$

where $a^{n+1} = a_y^3 a_x^2$ for $n = 4$ (step-edge diffusion case) and $a^{n+1} = a_y a_x^2$ for $n = 2$ (attachment/detachment case). Using the high-temperature approximation for step stiffness on a triangular lattice [19, 20],

$$\tilde{\beta} = \frac{2kT}{3a}\exp(\varepsilon_k/kT), \qquad (7)$$

and the reasonable estimate of the kink energy $\varepsilon_k \sim 0.1$ eV ([21], see above), we can generate an estimate of the step stiffness that provides a good description of the temperature dependence. Calculated values are shown in Table 2. The absolute magnitude of quantities calculated using this value of the kink energy may not be accurate, but for the purposes of this analysis will provide an internally consistent basis for discussion of the temperature dependence. Combining the calculated step stiffness with the temperature dependence of $G_1$ extracted from the fit shown in Fig. 5, the temperature dependence of the step-edge diffusion time constant $\tau_h$ is extracted, as shown in the third column of Table 2. The time constant decreases by approximately three orders of magnitude over the temperature range, consistent with an activated process.

The surprising results from the analysis of the autocorrelation functions are the strong dependence of the mean squared width on temperature (equivalent to the temperature dependence of the time correlation function), and the weak dependence of the correlation time on temperature. From Eq. 3a, we would expect that the correlation time would vary almost as strongly with temperature as the step-edge diffusion time constant $\tau_h$. Instead $\tau_c$ decreases by at most a factor of two over the same temperature range where $\tau_h$ decreases by a factor of a thousand. However, if we combine the expression relating the step width to the correlation length (for non-interacting steps, $w^2 = kTL/12\tilde{\beta}$) with the expected relationship between correlation time and correlation length, $\tau_c = kT(L/2\pi)^n/\Gamma_n\tilde{\beta}$, it is easy to show that:

$$G(t) = \frac{12w^2}{\pi^2}\Gamma\left(\frac{n-1}{n}\right)\left(\frac{t}{\tau_c}\right)^{1/n}. \tag{8}$$

Thus the observed parallel temperature dependence (Fig. 4) of the magnitude of G(t) and the values of the width require that the correlation time $\tau_c$ be constant (or weakly temperature dependent, given the substantial error bars), as observed via direct measurement of the autocorrelation function. This is a non-intuitive result if one interprets the correlation length as a fixed physical system size: in that case the relationship between correlation time and the underlying physical mobility $\Gamma$, $\tau_c = kT(L/2\pi)^n/\Gamma_n\tilde{\beta}$, would require a strong temperature dependence of the correlation time. Since this is not the case, we must conclude that the correlation length is a strongly temperature dependent quantity, and NOT a fixed physical limit on the effective experimental system size. This is indeed observed by calculating the effective length $L_{eff}$ from $w^2$ using Eq. 2, and the effective length $L_c$ from the correlation time, as shown in Table 2.  Here we see that $L_{eff}$ has the same temperature dependence as $L_c$ and the ratio between them is close to 1.

The temperature dependence of the effective system size can be understood in terms of the time dependence of the different wavelengths of step fluctuations shown in Eq. 5.  Because we measure over a finite time, we can only sample those wavelengths whose time constants are substantially smaller than the total measurement time.  To test this hypothesis, we consider a simple relationship in which we assume that some number, A, of decay times of a fluctuation are required to contribute coherently to the average.  We then relate the time constant $\tau_{q-min}$ of the maximum wavelength (minimum wavevector) contributing to the measurement via: $A\tau_{q-min} = t_m$.  Using Eq. 5, this yields an effective maximum wavelength (minimum $q$ in Eq. 4) that can contribute to our analysis of the correlation functions:

$$\lambda_{max} = 2\pi \left( \frac{a^{n+1} \tilde{\beta}}{\tau kT} \frac{t_m}{A} \right)^{1/n}. \qquad (9)$$

The value of A can be determined by equating $\lambda_{max}$ with the values of the length $L_c$ found from the correlation time, which is equivalent to equating $\tau_{q-min}$ to $\tau_c$, with the results shown in Table 2. We have found an average value for A of 9±3.  Repeating the calculation using the value of $L_{eff}$ yields indeterminate results, as the propagated error is as large as the calculated values of A (which are in the range of 25-35).  However, in both cases the assumption that the measurement time limits the effective correlation length is clearly consistent with the experimental temperature dependence.

We have also tested this hypothesis on previously measured data for an aluminum covered Si(111) surface [7, 8].  In that case, we also found that the measured correlation time was essentially constant [14] over a temperature range where the underlying physical time constant changes by three orders of magnitude, as shown in Table 3. Again, we used the measured values of the correlation time to calculate a correlation length $L_c$, set that equal to the maximum wavelength in Eq. 9 and extracted an average value of A = ~8. In this case also, the ratio of the measured effective lengths, $L_{eff}/L_c$ is approximately 1.

On the basis of this experimental evidence, it seems reasonable to conclude that, when the perfection of the physical system exceeds the wavelengths accessible in the measurement time, the measured correlation lengths are in fact a reflection of the experimental measurement time. This can be demonstrated theoretically, by paying careful attention to the details of the measurement. The key point to realize is that the measurement does not have a fixed "true" reference point of the average step position, as theoretical calculations normally do. Instead, the average step position is estimated as the average of the step positions over the measurement time interval $t_o \leq t \leq t_o + t_m$:

$$\bar{x}(y,t_o,t_m) = \frac{1}{t_m} \int_{t_o}^{t_o+t_m} x(y,t)dt. \tag{10}$$

The time correlation function G(t) (Eq. 1) is unaffected by this experimental issue, because it is calculated from the difference of two positions. However, Eqs. 2 and 3 for the mean-squared width and the autocorrelation function use absolute step positions, with the implicit assumption that x=0 is the average step position. Derivation of the equations needed to correct for this is shown in Appendix B. The key result for the effective mean-squared width $w_m^2(t_m)$, for measurement time $t_m$ is:

$$w_m^2(t_m) = \frac{kT}{\pi\tilde{\beta}} \int_{q_{min}}^{q_{max}} \frac{dq}{q^2} f\left(\frac{\Gamma_n \tilde{\beta} q^n t_m}{kT}\right) \tag{11}$$

where the function f(z) is defined in detail in Appendix B. If one defines a sharp cut-off point $z_c$ for f(z) (i.e. if f(z)=0 for $z \leq z_c$, and f(z)=1 for $z > z_c$) Eq. 11 reduces to a simple form:

$$w_m^2(t_m) = \frac{kT}{\pi\tilde{\beta}} \int_{q_c(t_m)}^{q_{max}} \frac{dq}{q^2}, \quad \text{where} \quad q_c(t_m) = \left(\frac{z_c kT}{\Gamma_n \tilde{\beta} t_m}\right)^{1/n}. \quad (12)$$

The expression for $q_c$ is identical to Eq. 9, with $z_c = A$. To find an appropriate numerical value for $z_c$, we have carried out, as shown in Appendix B, numerical integrations of the Edwards–Wilkinson (EW) equation (n = 2) and conserved noise 4-th order Langevin equation corresponding to step-edge-diffusion case (n = 4) for a system size of L sites with periodic boundary conditions. The simulation was performed for finite measurement times, and effective values of the width $w_m^2(t_m)$ and correlation time $\tau_m(t_m)$ were extracted from fits to the effective autocorrelation function. Extracting the effective lengths $L_{eff}^{(1)}$ and $L_{eff}^{(2)}$ from $w^2$ and $\tau_c$ respectively, as in the experimental data analysis, we find that their ratio is $L_{eff}(w_m^2)/L_{eff}(\tau_m) = 1.57$ for n = 2 and 0.95 for n = 4, as shown in Fig. 7. Both results are within reasonable statistical agreement with the measured ratios shown in Tables 2 and 3.

A second prediction from Eq. 12 is that the mean squared width will evolve as the 1/4 power of the measurement time. We have evaluated this possibility by breaking the total measurement time (26.9 s or 39.3 s) into small time intervals and evaluating $\bar{x}_m$ for each time interval. Then the standard deviation of the measured position values x around that average was evaluated over the time interval. The results are shown in Fig. 8 for the same data sets shown in the G(t) analysis in Fig. 3. The mean square width shows a strong time dependence that can be fit to a power law, although the fits are not as robust with respect to the exponent as the G(t) fits. The difference in the quality of the fits is not surprising as the G(t) analysis is a far better experimental design for the realities of the measurement than either the mean squared width or the autocorrelation function. In the cases shown, the fit exponents range from 0.20 to 0.25 with the shortest time data points (which have the worst statistics for evaluating $\bar{x}_m$) eliminated from the fits. The power-law dependence of the measured mean squared width on measurement time is direct evidence that the measured system size is an effective size determined by the measurement time.

**Conclusions**

We have evaluated the significance of system size for experimental measurements of fluctuating steps. The results show that the measurement technique itself can define an effective system size, which may be substantially smaller than the real spatial limitations of the physical system. As shown here, in the case of a temporal measurement of a point position, the effective size is determined by the total time over which the measurement is performed. That time is related to an effective system size through the time scales for relaxation of different wavelengths. Our experimental results are in good agreement with calculations in showing that the ratio of the measurement time and the characteristic decay time of a wavelength must exceed some characteristic value for that wavelength to be represented in the correlation functions. The characteristic value is theoretically predicted to be dependent both on the fluctuation mechanism, and the correlation function being measured. Experimental uncertainties prohibit a stringent test of the quantitative predictions, however for measurement of the autocorrelation function for conserved noise (step edge diffusion, $n = 4$), the characteristic value is ~9, with a distinct tendency for the value to increase with temperature. For non-conserved noise (EW case, $n = 2$), the characteristic value is ~8 for the autocorrelation function.

As one considers the consequences of structural fluctuations in nanoscale systems, these results indicate the necessity of including the time-scale of interest in the evaluation. Analytical results which involve integration over an infinite range of wavelengths, or the use of a fixed system size will give incorrect interpretation. Instead, the wavelengths available, given the physical time constants and the total observation time, must be considered properly.

**Acknowledgment**

This work has been supported by the NSF-NIRT under grant DMR-0102950 with partial support from U.S. Department of Energy Award No. DOE-FG02-01ER45939 and from the US-ONR. We also gratefully acknowledge support and SEF support from the NSF MRSEC under grant DMR 00-80008.

Table I: Temperature dependent values for step fluctuations on Ag(111). The total measurement time for each trajectory, x(t), used in the analysis is shown as $t_m$. $1/n$ is the exponent of the fit to $t^{1/n}$ behavior for the time correlation function (Eq. 1), $G_1$ is the magnitude of the time correlation function (Eq. 1), $w^2$ is the measured mean square width (Eq. 2).

| T, °C | $t_m$ (s) | $1/n$ | $G_{1exp}$, Å$^2$ | $w^2$, Å$^2$ | $\tau_c$, sec |
|---|---|---|---|---|---|
| 25 | 39.3 | 0.21±0.03 | 7.0±1.2 | 5.2±1.6 | 7.8±3.2 |
| 27 | 26.9 | 0.25±0.05 | 14.7±5.1 | 10.7±4.5 | |
| 43 | 26.9 | 0.20±0.02 | 16.0±2.0 | 10.6±1.9 | |
| 52 | 26.9 | 0.25±0.07 | 17.3±9.0 | 12.0±6.4 | |
| 63 | 26.9 | 0.23±0.05 | 12.8±4.3 | 9.2±5.0 | |
| 67 | 39.3 | 0.19±0.03 | 60±11 | 43±14 | 5.3±1.1 |
| 76 | 26.9 | 0.28±0.03 | 31±6 | 23±5 | |
| 108 | 26.9 | 0.28±0.04 | 29±7 | 20±8 | |
| 124 | 39.3 | 0.28±0.20 | 65±90 | 51±60 | 4.6±2.3 |
| 140 | 26.9 | 0.29±0.04 | 109±37 | 75±22 | |
| 177 | 39.3 | 0.22±.03 | 143±22 | 91±26 | 3.2±0.6 |

Table 2: Comparison of calculated values of the step properties. The step stiffness is calculated using Eq. 8 and a reasonable value of 0.1 eV for the kink energy. The edge-diffusion time constant $\tau_h$ is calculated using Eq. 1 and the fit parameters extracted from Fig. 4. Propagated errors on the calculated time constant are large, ranging from 40% at room temperature to nearly 90% at 177°C. $L_c$ is calculated from $\tau_c$ using Eq. 3a, and $L_{eff}$ is calculated from $w^2$ using Eq. 2. The values of A are calculated by equating $L_c$ to $\lambda_{max}$ in Eq. 9, and using the value of the measurement time from Table 1.

| T, °C | $\tilde{\beta}$, eV/Å | $\tau_h$, s | $L_c$, Å | $L_{eff}$, Å | $L_{eff}/L_c$ | $A(L_c)$ |
|---|---|---|---|---|---|---|
| 25 | 0.31 | 2.99E-06 | 1140±160 | 720±220 | 0.64±0.21 | 5.1±2.1 |
| 27 | 0.30 | 2.6E-06 | | 1550±650 | | ? |
| 43 | 0.26 | 9.02E-07 | - | 1160±210 | | - |
| 52 | 0.24 | 5.21E-07 | - | 1150±610 | | - |
| 63 | 0.22 | 2.78E-07 | | 960±520 | | |
| 67 | 0.22 | 2.23E-07 | 5480±1030 | 3770±1230 | 0.69±0.25 | 7.4±1.5 |
| 76 | 0.21 | 1.39E-07 | - | 1720±370 | | - |
| 108 | 0.17 | 3.07E-08 | - | 1180±470 | | - |
| 124 | 0.16 | 1.58E-08 | 3300±840 | 3300±3900 | 0.99±1.2 | 8.6±4.2 |
| 140 | 0.15 | 8.6E-09 | - | 3630±1100 | | - |
| 177 | 0.13 | 2.47E-09 | 5000±1560 | 3550±1010 | 0.71±0.30 | 12±2.3 |
| Theory | | | | | 0.95 | |

Table 3: Comparison of the values of the effective correlation lengths, $L_c$ is calculated from $\tau_c$ from previous measurements [7-9, 14, 25] using Eq. 3a, and $L_{eff}$ is calculated from $w^2$ using Eq. 2. The value of A is calculated using A = $t_m/\tau_c$, with $t_m$ = 38s.

| T, K | $\tau_a$, ms | $\tau_c$ | $L_c$, Å | $W^2$, Å$^2$ | $L_{eff}$, Å | A* | Leff/Lc |
|---|---|---|---|---|---|---|---|
| 770 | 260 | - - - | - - - | 74±59 | 2000±1600 | - - - | - - - |
| 870 | 29 | - - - | - - - | 91±37 | 1580±640 | - - - | - - - |
| 970 | 1.2 | 14±8 | 5800±1700 | 392±142 | 5300±1900 | 3±2 | 0.9±0.4 |
| 1020 | 0.28 | 3.2±2.1 | 5400±1800 | 429±147 | 5100±1800 | 12±8 | 1.0±0.4 |
| Theory | | | | | | | 1.57 |

**Figure captions**

Fig.1: Left panel – LEED pattern from Ag film on mica, showing epitaxial orientation of crystal structure with (111) surface. Right panel – 1000 nm by 1000 nm STM image of 100 nm Ag film. Tunneling conditions are: U=0.3 V, $I_t$= 0.1 nA.

Fig. 2: Pseudo image of Ag(111) at 320K obtained by scanning the STM tip repeatedly across a single line perpendicular to the step edge. Vertical axis shows 512 line scans initiated at equal sampling intervals over 26.9 s.

Fig. 3: (color on line) Time correlation functions for step fluctuations on Ag films at T = 52$^o$C, 67$^o$C, 140$^o$C and 177$^o$C. Each curve is the average of the time correlation functions obtained from ~20 independent measurements of step-position vs. time, standard deviations on the magnitudes of G(t) are around 15-30%, as listed in Table 1. The solid lines on the plots indicate power law scaling with exponents close to 1/4, as listed in Table I.

Fig. 4: (color on line) Plotting the time correlation function ($G_1$) vs. the mean squared width ($w^2$) demonstrates the linear relationship between the quantities. Standard deviations on the horizontal access have a similar proportionality to magnitude as on the vertical axis, and were included in the weighted fit yielding the best straight line fit (dashed line in the figure). The value of the slope is 1.4 ± 0.2.

Fig. 5 Magnitude of the time correlation function $G_1$ plotted in Arhennius form. The slope yields an apparent activation energy of 0.21 ± 0.02 eV.

Fig. 6: (color on line)Autocorrelation functions calculated from measured x(t) data using Eq. 3. The poorer statistics at long times make analysis difficult, and only four temperatures yielded interpretable results. Shown are data for T= 298K and 450K. Acceptable fits could not be obtained using the long time limit of Eq. 3, suggesting that the correlation time is within the time regime of the data shown. This is confirmed by fits to the full functional form (appendix A, Eq. 14), shown as the solid curves, which yields values of $\tau_c$ = 7.8 s and 3.2 s for the data shown.

Figure 7: Effective values of the length as a function of measurement time, determined from a numerical integration of the Edwards-Wilkinson equation for a system size L = 200 and the 4-th order Langevin equation for L=200. The straight lines are fits to a power law slope of $1/n = 0.5$ for the EW case and 0.25 for the 4-th order case.

Fig. 8 (color on line) The value of the mean squared width, calculated for the same data sets shown in Fig 3, for varying measurement times. Values are obtained by taking subsets of the total data set and calculating $\bar{x}$ and $w^2$ for each time subset. Fits in power law form can be obtained, but have considerable sensitivity to the data range of the fit (unlike the G(t) fits, which are robust). The solid curves shown have exponents $1/n = 0.25 \pm 0.03$ for T=52°C, $1/n = 0.20 \pm 0.02$ for T =67°C, $1/n = 0.25 \pm 0.02$ for 140°C, and $1/n = 0.25 \pm 0.02$ for 177 °C. All fits were forced to include (0,0).

# I. APPENDIX A: CALCULATION OF THE SHORT–TIME AND LONG–TIME BEHAVIORS OF $C(t)$ AND $G(t)$

We have the following definitions for the equilibrium time correlation function $G(t)$, squared interface width $w^2$ and autocorrelation function $C(t)$, respectively:

$$G(t) = \langle [x(t) - x(0)]^2 \rangle, \tag{1}$$

$$w^2 = \langle [x(t) - \overline{x}]^2 \rangle, \tag{2}$$

$$C(t) = \langle [x(t) - \overline{x}][x(0) - \overline{x}] \rangle, \tag{3}$$

where $\overline{x}$ is the average step position. Two simple relations can be used for further simplifications:

$$w^2 = C(0) \tag{4}$$

and

$$G(t) = 2[w^2 - C(t)]. \tag{5}$$

Using a Langevin description of the dynamics (see Ref. [14]) and considering a finite system of size $L = Na$ where $a$ is the lattice parameter, it is straightforward to obtain the following expression for $C(t)$:

$$C(t) = \frac{2}{L} \sum_j \frac{K_1}{q_j^2} \exp(-K_2 q_j^n t), \tag{6}$$

where $K_1 = kT/\tilde{\beta}$ and $K_2 = \Gamma_n \tilde{\beta}/(k_B T)$, with $n = 2$, $\Gamma_n = \Gamma_a$ for non-conserved dynamics, and $n = 4$, $\Gamma_n = \Gamma_h$ for conserved dynamics. The sum over the integer $j$ in Eq. (6) runs from 1 to $N/2$ with $q_j = 2\pi j/L$.

Using this expression for $C(t)$, one obtains

$$w^2 = C(0) = \frac{K_1 L}{2\pi^2} \sum_{j=1}^{N/2} \frac{1}{j^2}. \tag{7}$$

Replacing the upper limit by infinity (for large $N$), the sum over $j$ is obtained as $\zeta(2) = \pi^2/6$, which leads to the result

$$w^2 = \frac{K_1 L}{12} = \frac{kTL}{12\tilde{\beta}}. \tag{8}$$

It is clear from Eq. (6) that at long times, the decay of $C(t)$ is governed by the mode with the smallest $j$ ($j = 1$). This leads to the following long–time asymptotic behavior of the

steady–state autocorrelation function:

$$C(t) = \frac{K_1 L}{2\pi^2} e^{-t/\tau_c} = \frac{kTL}{2\pi^2 \tilde{\beta}} e^{-t/\tau_c}, \qquad (9)$$

with the correlation time $\tau_c$ given by

$$\tau_c(L) = \frac{1}{K_2 q_1^n} = \frac{kTL^n}{(2\pi)^n \Gamma_n \tilde{\beta}}. \qquad (10)$$

Finally, the time correlation function $G(t)$ at relatively short times ($t \ll \tau_c$) is obtained as

$$G(t) = \frac{2\Gamma(1 - 1/n)}{\pi} \left(\frac{kT}{\tilde{\beta}}\right)^{1-1/n} \Gamma_n^{1/n} t^{1/n}, \qquad (11)$$

where $\Gamma$ is the Euler gamma function. Note that the numerical coefficient $2\Gamma[1 - 1/n]/\pi$ becomes 1.128 for $n = 2$ and 0.780 for $n = 4$. On the other hand, for $t \gg \tau_c$, $C(t)$ decays exponentially (see Eq. (9)), and the time correlation function becomes

$$G(t) = 2[C(0) - C(t)] = 2w^2 - \frac{K_1 L}{\pi^2} e^{-t/\tau_c}, \qquad (12)$$

so that $G(t)$ reaches a time independent saturation value of $2w^2$ at very long times.

In the present analysis, we have used these expressions for $w^2$, $C(t)$ and $G(t)$. Somewhat different expressions for $w^2$ and $C(t)$ are obtained if the discrete sum in Eq. (6) is replaced by an integral:

$$C(t) = \frac{K_1}{\pi} \int_{q_{min}}^{q_{max}} \frac{dq}{q^2} \exp(-K_2 q^n t), \qquad (13)$$

where $q_{min} = 2\pi/L$ and $q_{max} = \pi/a$. If the upper limit of the integral is replaced by infinity, then one can obtain a closed-form expression for $C(t)$:

$$C(t) = C(0) \left[\exp(-t/\tau_c) - \Gamma[1 - 1/n, t/\tau_c] \left(\frac{t}{\tau_c}\right)^{1/n}\right], \qquad (14)$$

where $C(0) = w^2 = kTL/(2\pi^2 \tilde{\beta})$, $\tau_c$ is the correlation time defined above, and $\Gamma[x, y]$ is the incomplete gamma function that reduces to the Euler gamma function for $y = 0$ (i.e., $\Gamma[x, 0] = \Gamma(x)$). In the short-time regime, $t \ll \tau_c$, Eq. (14) becomes

$$C(t) = \frac{kTL}{2\pi^2 \tilde{\beta}} \left[1 - \Gamma[1 - 1/n] \left(\frac{t}{\tau_c}\right)^{1/n}\right], \qquad (15)$$

which leads to the same expression as Eq. (11) for $G(t)$. In the long-time limit, $t \gg \tau_c$, this treatment leads to an asymptotic behavior which is *different* from that of Eq. (9):

$$C(t) = \frac{kTL}{2\pi^2 \tilde{\beta}} \frac{\tau_c}{nt} e^{-t/\tau_c}. \qquad (16)$$

The difference between the results for $w^2$ and $C(t)$ obtained using the sum and the integral arises due to the "infrared" divergence of the integral. The same expression for $G(t)$ is obtained in the two methods because this divergence does not occur in this case.

## II. APPENDIX B: CALCULATION AND SIMULATION OF THE EXPERIMENTALLY MEASURED AUTOCORRELATION FUNCTION

In the experiment, the step position $x(y, t)$ is measured at a particular point $y$ along the step over the time interval $t_0 \leq t \leq t_0 + t_m$, where the initial time $t_0$ is such that the step has reached equilibrium at temperature $T$ at this time, and $t_m$ is the measurement time. The average step position is then assumed to be

$$\bar{x}(y, t_0, t_m) \equiv \frac{1}{t_m} \int_{t_0}^{t_0+t_m} x(y, t) dt, \quad (17)$$

where we have included $t_0$ and $t_m$ in the argument of $\bar{x}$ because the average would, in general, depend on these two times. The experimentally measured autocorrelation function $C_m(t, t_m)$ is then defined in terms of the deviations of the instantaneous step position from the assumed average value. Specifically,

$$C_m(t, t_m) \equiv \frac{1}{t_m - t} \int_{t_0}^{t_0+t_m-t} \langle [x(y, t') - \bar{x}(y, t_0, t_m)][x(y, t' + t) - \bar{x}(y, t_0, t_m)] \rangle dt', \quad (18)$$

with $t < t_m$. Here $\langle \cdots \rangle$ represents an average over different realizations of the stochastic evolution of the step position. Note that $C_m$ does not depend on $t_0$ if the step is in the equilibrium state at this time. The right-hand side of Eq. (18) can be written in terms of integrals of the autocorrelation function of fluctuations of the step position (from its true average which is assumed to be zero) in the long-time equilibrium state. We assume that the stochastic dynamics of the step is governed by a linear Langevin equation of the type discussed in Ref. [14]. It is then possible to calculate this autocorrelation function exactly. As discussed in Ref. [14], the autocorrelation function of $\tilde{x}(q, t)$, the Fourier transform of $x(y, t)$, in the equilibrium state has the form

$$\langle \tilde{x}(q, t_1) \tilde{x}(-q, t_2) \rangle = \frac{K_1}{q^2} e^{-K_2 q^n |t_1 - t_2|}, \quad (19)$$

where $K_1$ and $K_2$ have been defined in Appendix A. Using this result, it is straightforward to derive the following expression for $C_m(t, t_m)$:

$$C_m(t,t_m) = \frac{K_1}{\pi} \int_{q_{min}}^{q_{max}} \frac{dq}{q^2} \left[ e^{-K_2 q^n t} + \frac{2}{(K_2 q^n t_m)^2}(K_2 q^n t_m - 1 + e^{-K_2 q^n t_m}) \right.$$
$$\left. - \frac{2}{K_2 q^n t_m}\left(2 - \frac{1}{t_m - t}\frac{1}{K_2 q^n}(1 - e^{-K_2 q^n (t_m - t)})(1 + e^{-K_2 q^n t})\right) \right], \quad (20)$$

where $q_{min}$ and $q_{max}$ have been defined in Appendix A. The right-hand side of Eq. (20) can be evaluated as a function of $t$ for different values of the measurement time $t_m$. We first consider the case $t = 0$, which yields the measured value of the step width:

$$w_m^2(t_m) \equiv C_m(t=0, t_m) = \frac{K_1}{\pi} \int_{q_{min}}^{q_{max}} \frac{dq}{q^2}$$
$$- \frac{K_1}{\pi} \int_{q_{min}}^{q_{max}} \frac{dq}{q^2} \frac{2}{(K_2 q^n t_m)^2}(K_2 q^n t_m - 1 + e^{-K_2 q^n t_m}). \quad (21)$$

In the right-hand side of Eq. (21), the first term represents $w^2$ where $w$ is the *true* width of the step, and the second term represents $\langle [\bar{x}(y, t_0, t_m)]^2 \rangle$, the mean squared value of the *apparent* average step position obtained from measurements over time $t_m$. Clearly, the value of $w_m^2$ is always smaller than $w^2$, and the difference depends on the measurement time $t_m$: the second term in the right-hand side of Eq. (21) goes to zero as $t_m \to \infty$, but the decay to zero is rather slow. To analyze the dependence of $w_m^2$ on $t_m$, it is convenient to write Eq. (21) as

$$w_m^2(t_m) = \frac{K_1}{\pi} \int_{q_{min}}^{q_{max}} \frac{dq}{q^2} f(K_2 q^n t_m) \quad (22)$$

where $f(z) \equiv [1 - 2(z - 1 + \exp(-z))/z^2]$ is a function that goes from 0 to 1 as the argument $z$ is increased from 0 to $\infty$. A simple approximation for the dependence of $w_m^2$ on $t_m$ may be obtained by assuming that $f(z) = 0$ for $z \leq z_c$ and $f(z) = 1$ for $z > z_c$, where $z_c \simeq 2.6$ is the value of $z$ for which $f(z) = 0.5$. Using this approximation for $f(z)$, Eq. (22) may be written as

$$w_m^2(t_m) = \frac{K_1}{\pi} \int_{q_c(t_m)}^{q_{max}} \frac{dq}{q^2}, \quad (23)$$

where the lower cutoff $q_c(t_m)$ of the $q$-integral is given by

$$q_c(t_m) = \left(\frac{z_c}{K_2 t_m}\right)^{1/n}. \quad (24)$$

Noting that the right-hand side of Eq. (23) represents the square of the equilibrium width of a step of length $2\pi/q_c$, we arrive at the result that the measured width $w_m$ corresponds to an *effective* step length $L_{eff}$ that depends on the measurement time $t_m$ as

$$L_{eff}(t_m) = \frac{2\pi}{q_c(t_m)} = 2\pi \left(\frac{K_2 t_m}{z_c}\right)^{1/n}. \qquad (25)$$

This is essentially the same as Eq. (9) with $A = z_c \simeq 2.6$.

It is interesting to note that the behavior of $w_m$ as a function of $t_m$ is qualitatively similar to the dependence of the step width $w$ on the "equilibration time" $t_{eq}$ that measures the time over which a step evolves starting from a completely straight initial configuration. The step width at time $t_{eq}$ is given by

$$w^2(t_{eq}) = \frac{K_1}{\pi} \int_{q_{min}}^{q_{max}} \frac{dq}{q^2}(1 - e^{-2K_2 q^n t_{eq}}). \qquad (26)$$

This is similar to Eq. (21), with the function $f(K_2 q^n t_m)$ replaced by the new function $g(K_2 q^n t_{eq})$ where $g(z) \equiv 1 - e^{-2z}$. The functions $f(z)$ and $g(z)$ have the same asymptotic values for $z = 0$ and $z \to \infty$, so that the approximate analysis described above may also be carried out for the dependence of $w^2(t_{eq})$ on $t_{eq}$. One then obtains a relation between the "effective length" $L_{eff}$ and the equilibration time $t_{eq}$ that is very similar to Eq. (25), but with a smaller value ($\sim 0.35$) of the parameter $z_c$.

We have studied the dependence of the rate of the temporal decay of the measured autocorrelation function $C_m(t, t_m)$ (i.e., the correlation time extracted from the measured autocorrelation function) on the measurement time $t_m$ by numerically evaluating the right-hand side of Eq. (20) (with the $q$-integral replaced by a sum over discrete $q$-values) for different choices of $t_m$. We have also simulated the behavior of $C_m(t, t_m)$ by carrying out numerical integrations of the Edwards–Wilkinson (EW) ($n = 2$) equation. The spatially discretized form of the equations used in the numerical work is

$$\frac{dx_i}{dt} = (x_{i-1} + x_{i+1} - 2x_i) + \eta_i(t), \qquad (27)$$

where $x_i(t)$ represents the step position at lattice site $i$ at time $t$, and $\eta_i(t)$ are uncorrelated random variables with zero mean and unit variance. For a system of $L$ sites with periodic boundary conditions, the equilibrium width is then given by $w^2(L) = L/24$, and the correlation time in the long-time equilibrium state is $\tau_c(L) = (L/2\pi)^2$. The simulations were done for $L = 200$, using $t = 10^4$ for equilibration (starting from a straight step profile), and the results were averaged over $10^4$ runs. The correlation function $C_m(t, t_m)$ was obtained by following the same procedure as that used in the experiments.

Fig. 1 shows the results for $C_m(t, t_m)$ obtained for $t_m = 400$. Both the data points obtained from the simulations and the analytic result obtained from Eq. (20) are shown in the plot. It is clear that the simulation results are in exact agreement with the analytic prediction. We have checked that $C_m$ obtained from experimental data has a behavior similar to that shown in Fig. 1, e.g., it becomes negative at large times. In the inset of Fig. 1 we show that the decay of $C_m(t, t_m)$ (with $t_m = 400$) for small values of $t$ is well-represented by an exponential form, $C_m(t, t_m) \sim \exp[-t/\tau_m(t_m)]$. The value of the "measured correlation time" $\tau_m$ may be extracted from exponential fits similar to the one shown in the plot. Note that the general form of the dependence of $C_m(t, t_m)$ on $t$ is very similar to that shown in Fig. 6.

Using the relations, $w_m^2(t_m) = L_{eff}^{(1)}(t_m)/24$ and $\tau_m(t_m) = [L_{eff}^{(2)}(t_m)/2\pi]^2$, we can extract two "effective length" parameters $L_{eff}^{(1)}$ and $L_{eff}^{(2)}$ from the results for $w_m^2$ and $\tau_m$. Note that $L_{eff}^{(1)} = L_{eff}$ and $L_{eff}^{(2)} = L_c$ in Tables 2 and 3. Fig. 7 shows the dependence of $L_{eff}^{(1)}$ and $L_{eff}^{(2)}$ on the measurement time $t_m$. In both cases, the data are well-described by a power-law with exponent $1/2$. From the fits shown in the plot, we get for the $n = 2$ case:

$$L_{eff}^{(1)}(t_m) \simeq 3.28 t_m^{1/2}, \tag{28}$$

and

$$L_{eff}^{(2)}(t_m) \simeq 2.09 t_m^{1/2}. \tag{29}$$

Comparison with Eq. (9), using $K_2 = \Gamma_n \tilde{\beta}/(kT) = 1$ and $n = 2$, yields the values $A \simeq 3.7$ and $A \simeq 9.0$, from the data for $w_m^2$ and $\tau_m$, respectively. Note that the value of $A$ extracted from the data for $w_m^2$ is consistent with the simple estimate mentioned above, and the value of $A$ obtained from the measured values of $\tau_m$ is close to that found experimentally.

On the other hand, for the conserved fourth-order Langevin equation ($n = 4$) case, a similar analysis with $L = 200$ yields the following results:

$$L_{eff}^{(1)}(t_m) \simeq 3.29 t_m^{1/4}, \tag{30}$$

and

$$L_{eff}^{(2)}(t_m) \simeq 3.45 t_m^{1/4}. \tag{31}$$

These quatities are shown in Fig. 7 as well. Comparison with Eq. (9), using $K_2 = \Gamma_n \tilde{\beta}/(kT) = 1$ and $n = 4$, yields the values $A \simeq 13.3$ and $A \simeq 11.0$, from the data for $w_m^2$ and $\tau_m$, respectively, in good agreement with the value found experimentally.

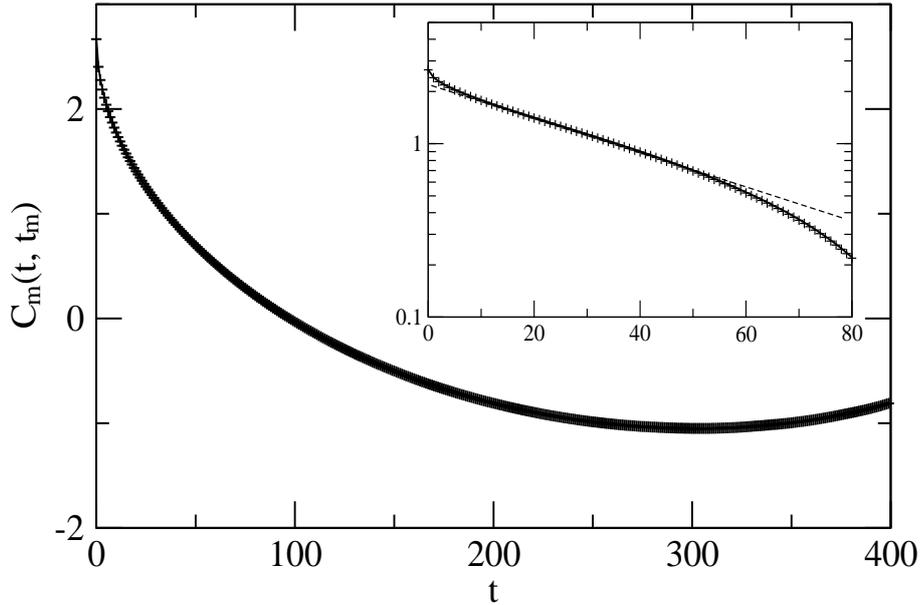

FIG. 1: Analytic and simulation results for the correlation function $C_m(t, t_m)$ for the Edwards-Wilkinson equation ($n = 2$). Plus signs: Simulation data ($L = 200$, $t_m = 400$) for $C_m$ as a function of $t$. Solid line: analytic results for the same quantity. Inset: same as the main figure, but using semi-log scale and $t \leq 80$; the dashed line represents an exponential fit to the short-time data for $C_m(t, t_m)$.

Finally, we note that the experimentally obtained values of the time correlation function

$$G(t) = \langle [x(y, t_0 + t) - x(y, t_0)]^2 \rangle \qquad (32)$$

are not affected by the measurement time $t_m$ because measurements of this quantity do not require estimating the value of the average step position from the data for $x(y, t)$. This is confirmed by our simulation data.

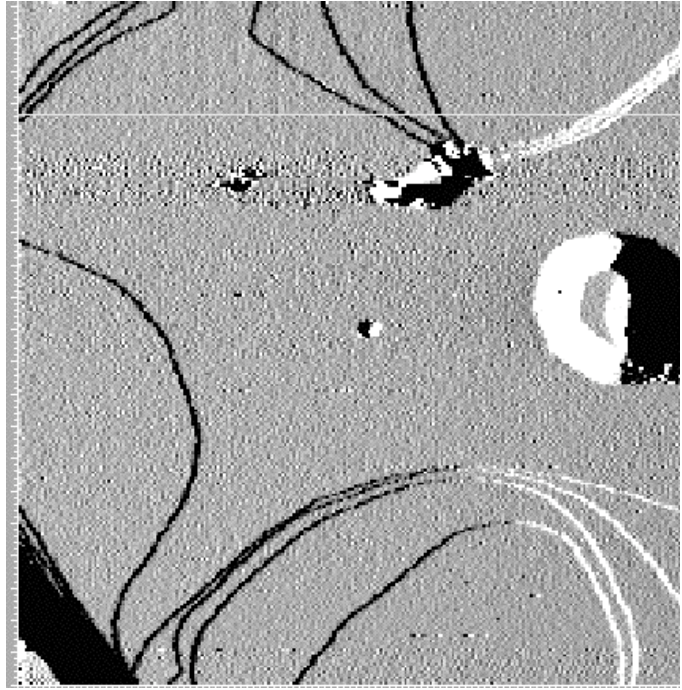

Fig.1: 200 nm x 200 nm STM image of 100 nm Ag film. Tunneling conditions are: U=0.3 V, $I_t$= 0.1 nA.

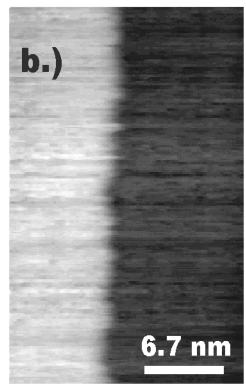

Fig. 2: Pseudo image of Ag(111) at 320K obtained by scanning the STM tip repeatedly across a single line perpendicular to the step edge. Vertical axis shows 512 line scans initiated at equal sampling intervals over 26.9 s.

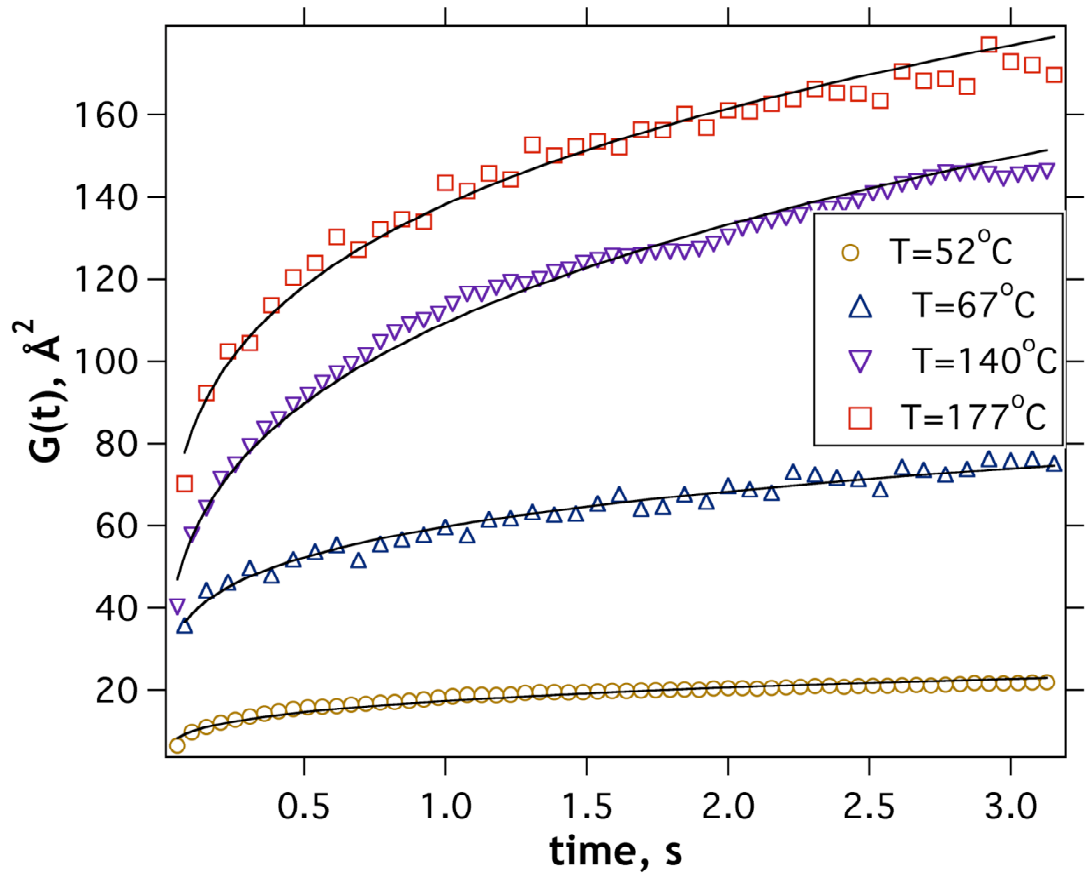

Fig. 3: (color online) Time correlation functions for step fluctuations on Ag films at T = 52°C, 67°C, 140°C and 177°C. Each curve is the average of the time correlation functions obtained from ~20 independent measurements of step-position vs. time, standard deviations on the magnitudes of G(t) are around 15-30%, as listed in Table 1. The solid lines on the plots indicate power law scaling with exponents close to 1/4, as listed in Table I.

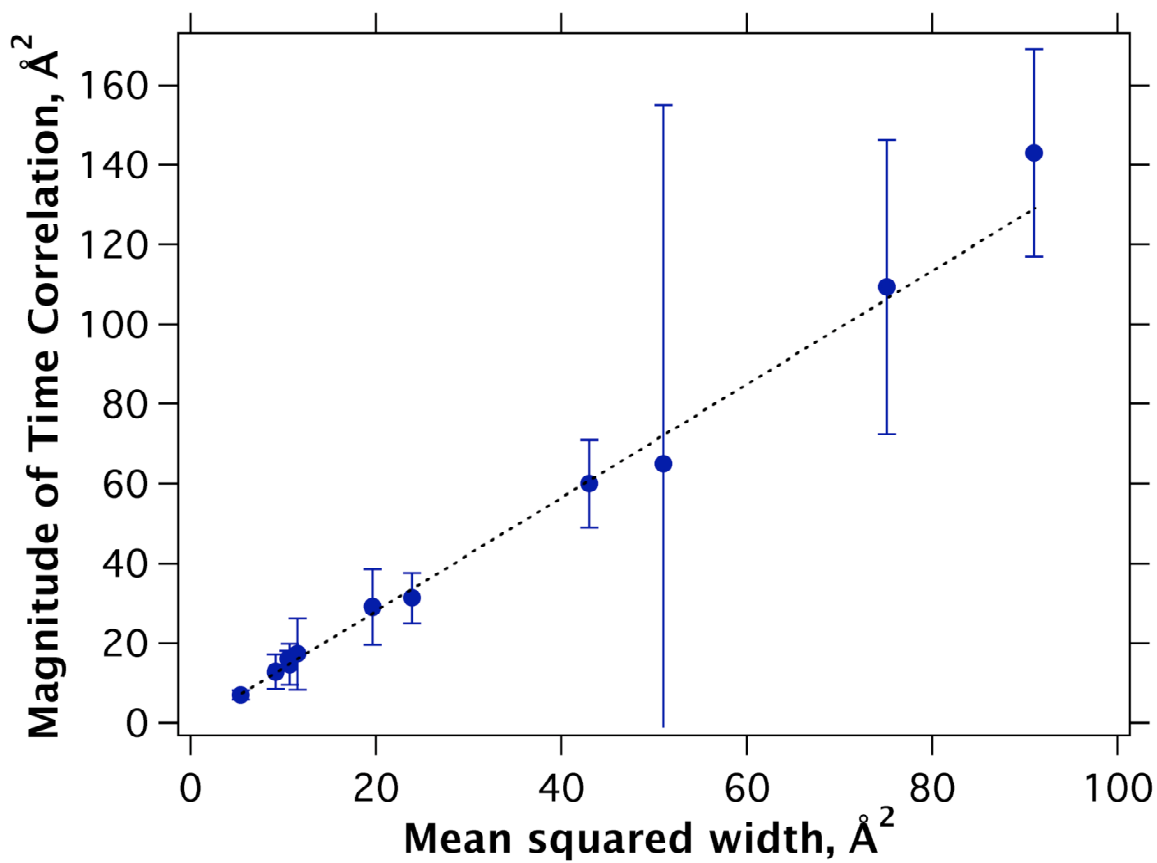

Fig. 4: Plotting the time correlation function ($G_1$) vs. the mean squared width ($w^2$) demonstrates the linear relationship between the quantities. Standard deviations on the horizontal access have a similar proportionality to magnitude as on the vertical axis, and were included in the weighted fit yielding the best straight line fit (dashed line in the figure). The value of the slope is 1.4 ± 0.2.

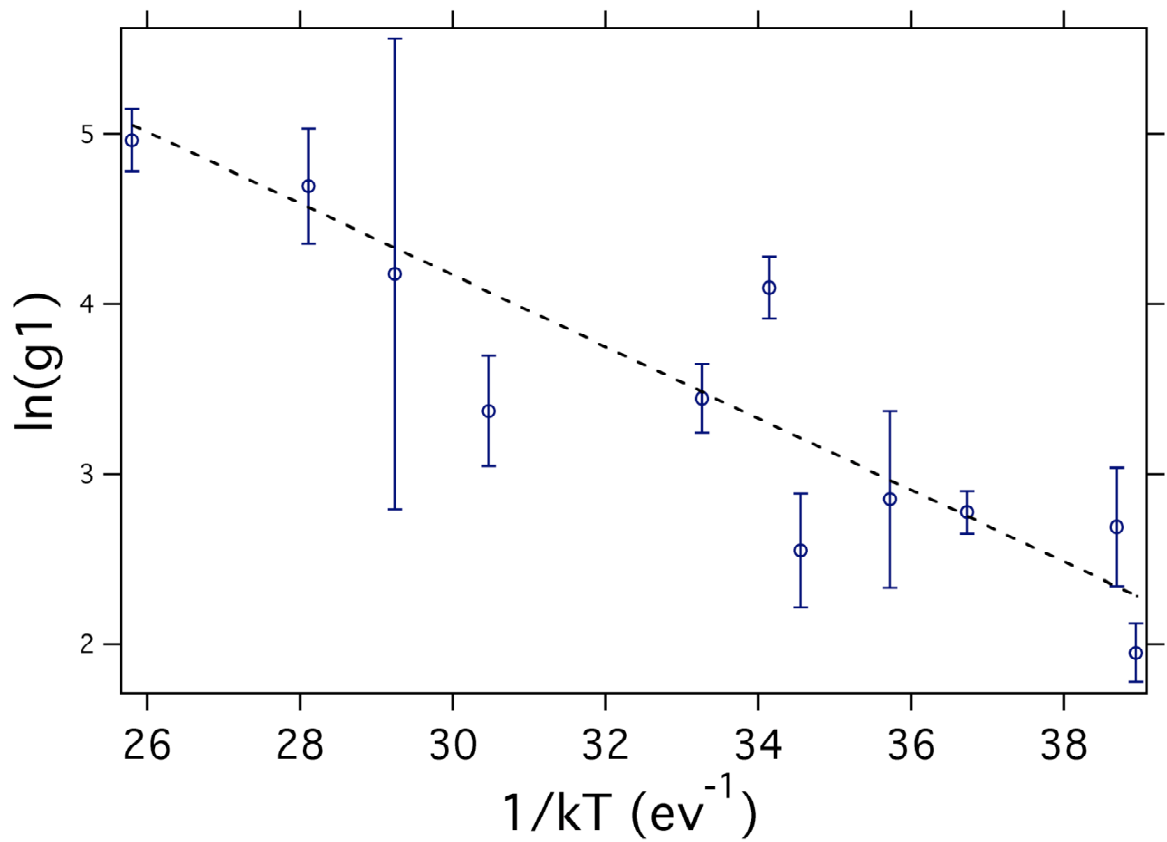

Fig. 5 Magnitude of the time correlation function $G_1$ plotted in Arhennius form. The slope yields an apparent activation energy of $0.21 \pm 0.02$ eV.

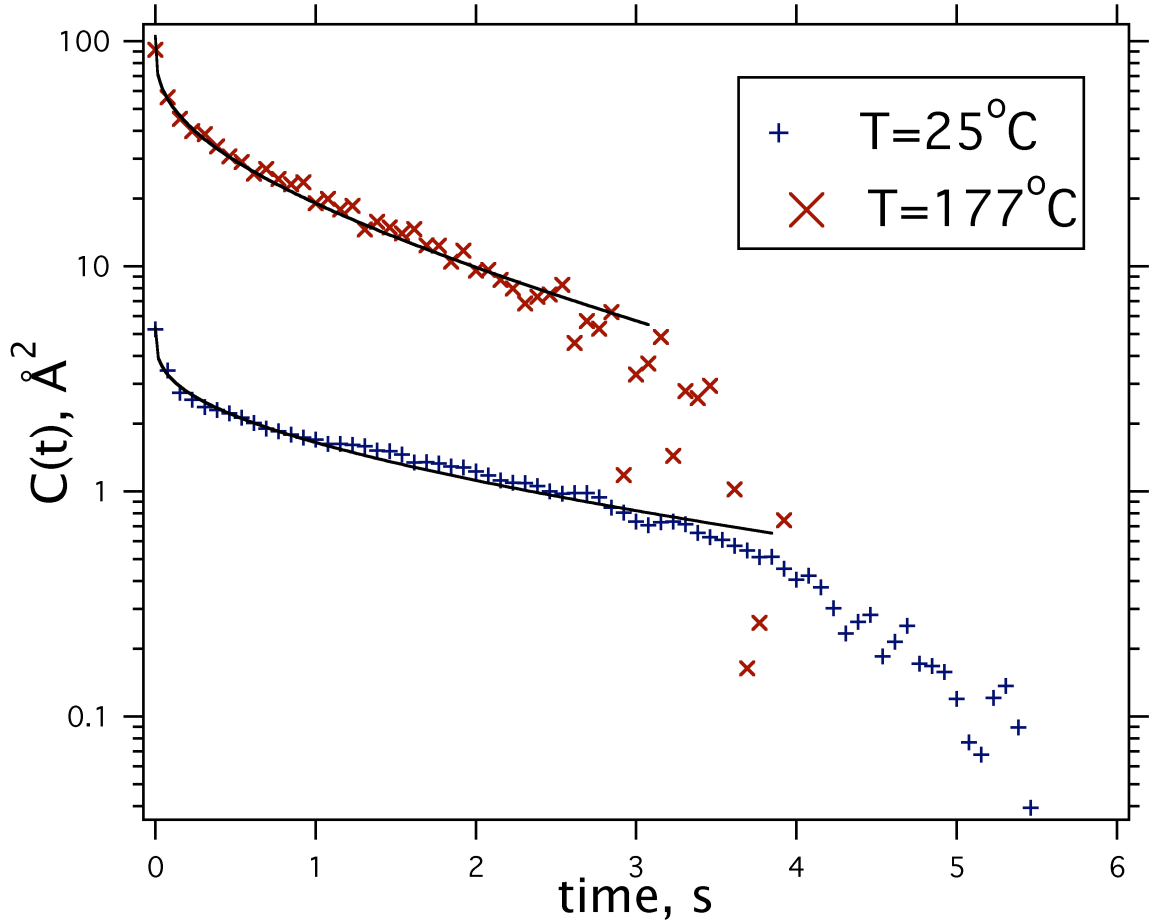

Fig. 6: (color online) Autocorrelation functions calculated from measured x(t) data using Eq. 3. The poorer statistics at long times make analysis difficult, and only four temperatures yielded interpretable results. Shown are data for T= 298K and 450K. Acceptable fits could not be obtained using the long time limit of Eq. 3, suggesting that the correlation time is within the time regime of the data shown. This is confirmed by fits to the full functional form (appendix A, Eq. 14), shown as the solid curves, which yields values of $\tau_c$ = 7.8 s and 3.2 s for the data shown.

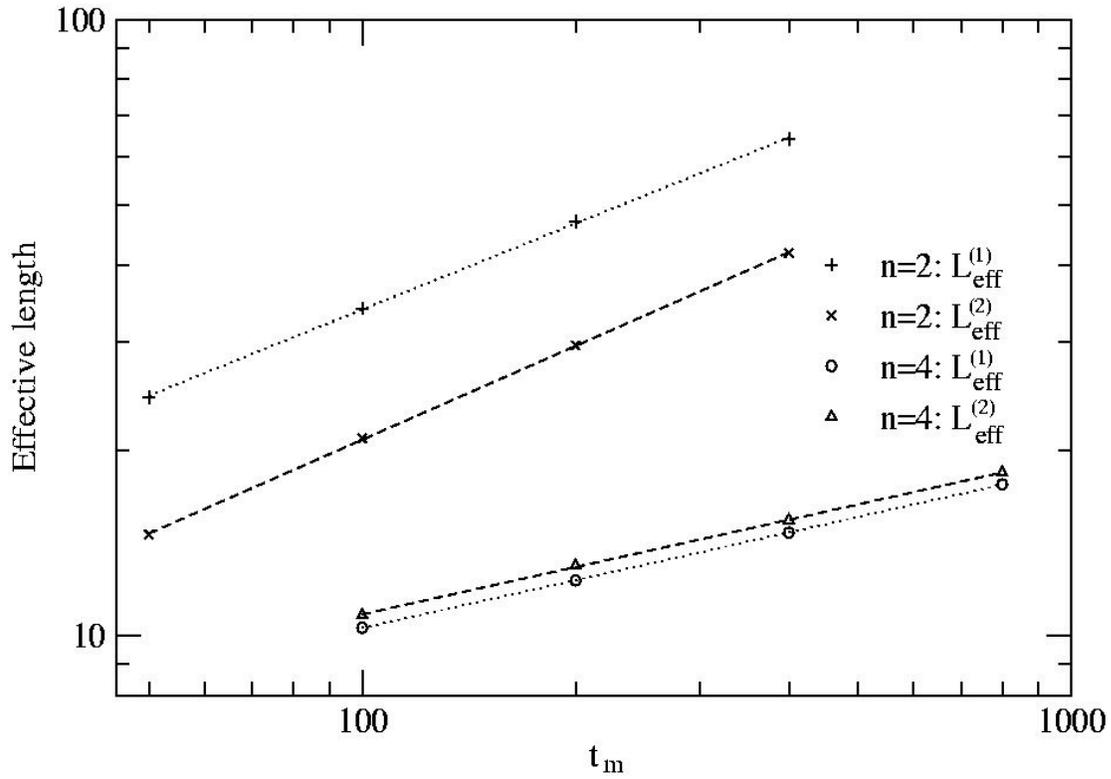

Figure 7: Effective values of the length as a function of measurement time, determined from a numerical integration of the Edwards-Wilkinson equation for a system size L = 200 and the 4-th order Langevin equation for L=200. The straight lines are fits to a power law slope of $1/n = 0.5$ for the EW case and 0.25 for the 4-th order case.

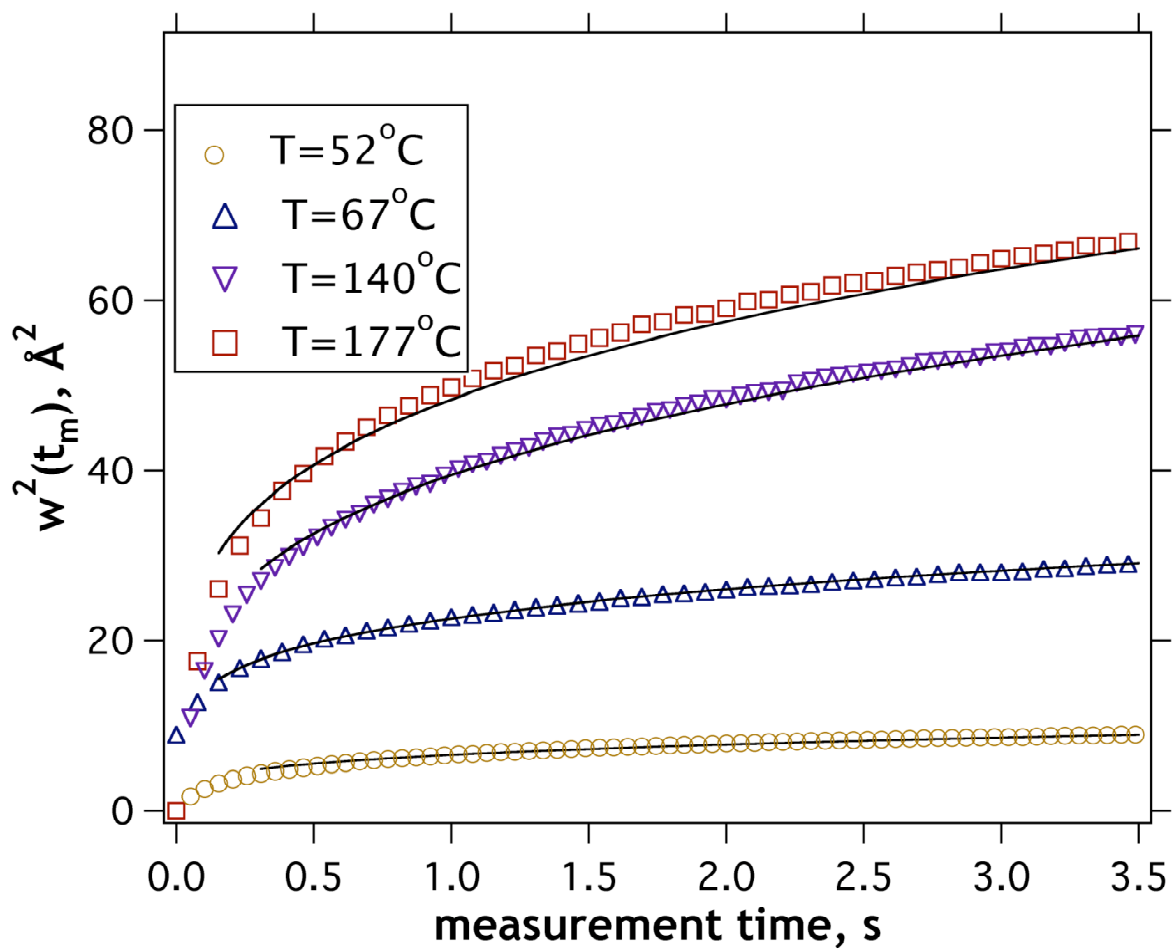

Fig. 8 (color online) The value of the mean squared width, calculated for the same data sets shown in Fig 3, for varying measurement times. Values are obtained by taking subsets of the total data set and calculating $\bar{x}$ and $w^2$ for each time subset. Fits in power law form can be obtained, but have considerable sensitivity to the data range of the fit (unlike the G(t) fits, which are robust). The solid curves shown have exponents $1/n = 0.25\pm0.03$ for T=52°C, $1/n = 0.20 \pm0.02$ for T =67°C, $1/n = 0.25\pm0.02$ for 140°C, and $1/n = 0.25\pm0.02$ for 177 °C. All fits were forced to include (0,0).